\documentclass[a4paper,dvips 2t,nofootinbib,notitlepage]{revtex4-1}
\usepackage{setspace}
\usepackage{epsfig,graphics,amsmath,amssymb}
\usepackage{rotate}
\usepackage{latexsym}
\usepackage{graphicx}

\newcommand{\mathsym}[1]{{}}
\newcommand{\unicode}[1]{{}}

\begin{document}
\title{\bf Constraining unparticle physics from $CP$ violation in Cabibbo-favored decays of D mesons}

\author{ Mohammadmahdi Ettefaghi}\email{mettefaghi@qom.ac.ir} \author{Reza Moazzemi} \email{r.moazzemi@qom.ac.ir}
\author{ Mohsen Rousta}

\affiliation{Department of Physics, The University of Qom, Ghadir Blvd., Qom 371614-611, I. R. Iran}

\begin{abstract}
According to the standard model, the Cabibbo-favored (CF) decays are $CP$ conserve at tree level. Observation of any finite $CP$ asymmetry can be received as a signal of new physics. In CF charm meson decays, $ D^0 \rightarrow K^- \pi^+ $ and $ D^+ \rightarrow K_s^0 \pi^+ $, the following experimental values for their $CP$ asymmetry are reported, respectively: ($0.3  \pm  0.7$) $\%$ and ($-0.41  \pm  0.09$) $\%$. The value of the later can be attributed to the mixing of $ K^0 $ and $ \overline{K^0} $, however, its contribution is about ($-0.332 \pm 0.006 $) $\%$.  In this paper, we use these experimental results to constrain the unparticle stuff as a new physics which may contribute to these $CP$ asymmetries.

\end{abstract}

\maketitle

\section{Introduction}

In the standard model (SM), $CP$ violation comes from the complex valued nature of the Cabibbo-Kobayashi-Maskawa (CKM) matrix, and in fact from a residual imaginary phase \cite{kobayashi1973cp}.  There are two types of $CP$ violation: direct $CP$ violation ($CP$ violation in decay) and indirect $CP$ violation ($CP$ violation with mixing). In charged mesons (such as $ D^+ $), since there is no mixing with their antiparticles, only direct $CP$ violation observes. The direct $CP$ asymmetry for typical $ D \to f $ decay is defined as:

\begin{eqnarray}\label{cp}
A_{CP}=\dfrac{\Gamma (D \to f)-\Gamma (\bar{D} \to \bar{f})}{\Gamma (D \to f)+\Gamma (\bar{D} \to \bar{f})} = \dfrac{\vert A (D \to f) \vert^2 -\vert A (\bar{D} \to \bar{f}) \vert^2}{\vert A (D \to f) \vert^2 -\vert A (\bar{D} \to \bar{f}) \vert^2},
\end{eqnarray}
where $ \Gamma $ and $ A $ are the partial decay width and  decay amplitude, respectively. To have a $CP$ violation, we need two amplitudes, $A_1$ and $A_2$, with different $CP$-conserved phase and also different $CP$-violated phase. Rewriting $A (\bar{D} \to \bar{f})$ as $\bar{A}$  and defining $ A_i = \vert A_i \vert e^{i(\phi_{i}+\delta_{i})} $ and $ \bar{A_i} = \vert A_i \vert e^{i(-\phi_{i}+\delta_{i})} $ , Eq. (\ref{cp}) can be written as

\begin{eqnarray}\label{e2}
A_{CP}=\dfrac{\vert A \vert^2 - \vert \bar{A} \vert^2}{\vert A \vert^2 + \vert \bar{A} \vert^2} = \dfrac{2 \vert A_1\vert \vert A_2\vert \sin \Delta\phi \sin \Delta\delta }{\vert A_1\vert^2 + \vert A_2\vert^2 + \vert A_1 A_2 \vert \cos \Delta\phi \cos \Delta\delta} \, .
\end{eqnarray}
This equation also confirms that to have a nonvanishing $CP$ violation in decay, two amplitudes with two nonzero $CP$-conserved and -violated phase  differences are needed. For latter uses we introduce the ratio of $A_1$ and $A_2$ by $r_f=A_1/A_2$.

Historically, $CP$ violation discovered and observed in 1964 for K mesons by Cronin and Fitch \cite{christenson1964evidence}. Then, it observed in many decays for B mesons (see for instance\cite{carter1981cp,abe2001observation}). But in the standard model, for D mesons, $CP$ violation is predicted to be very
small, $\leq\mathcal{O}(0.1 \%)$  \cite{grossman2007new}. In fact, this issue is also obvious from the CKM matrix which is, up to order $\lambda^6$, written as

\begin{eqnarray}
V_{CKM}=
\begin{pmatrix}
V_{ud} & V_{us} & V_{ub} \\
V_{cd} & V_{cs} & V_{cb} \\
V_{td} & V_{ts} & V_{tb}
\end{pmatrix}\nonumber
\end{eqnarray}
\begin{eqnarray}\label{ckm}
\small{
=\begin{pmatrix}
1-\dfrac{1}{2}\lambda^2-\dfrac{1}{8}\lambda^4 & \lambda & A\lambda^3(\rho -i\eta) \\
-\lambda+\dfrac{1}{2}A^2\lambda^5[1-2(\rho +i\eta)] & 1-\dfrac{1}{2}\lambda^2-\dfrac{1}{8}\lambda^4(1+4A^2) & A\lambda^2 \\
 A\lambda^3[1-(\rho +i\eta)(1-\dfrac{1}{2}\lambda^2)]& -A\lambda^2+\dfrac{1}{2}A\lambda^4[1-2(\rho +i\eta)] & 1-\dfrac{1}{2}A^2\lambda^4
\end{pmatrix}+ \mathcal{O} (\lambda^6)  } \, .
\nonumber \\
\end{eqnarray}
Since in D mesons we deal only with the first and second generations, the imaginary part of the relevant elements in the above matrix are of order $\lambda^5$.
There was also no experimental observation of $CP$ violation for D mesons till 2012 when LHCb reported a $CP$ asymmetry for this meson's family \cite{fringscp}, though the recent LHCb measurement shows no evidence for $CP$ violation \cite{lhcb2016}. However, the large uncertainties which exist in these measurements allow one to assume a new physics (NP) beyond the SM for explaining a possible deviation from SM and(or) putting some constrains on the parameter space of such NP. 

Unparticle is a subjective theory that Georgi introduced in 2007 \cite{georgi2007unparticle}. In addition to the experimental searches  \cite{skh,vkh,vkh2,ams}, unparticle physics is  widely considered in various topics of high energy physics, such as in cosmology \cite{davoudiasl2007constraining,lewis2007cosmological,mcdonald2009cosmological}, astronomy \cite{freitas2007astro,hannestad2007unparticle,deshpande2008long}, neutrino oscillation \cite{boyanovsky2010oscillation} and even in solid state and atomic physics \cite{pwphil,ali,klimph,jpfl,klim,akar,mfwon} etc.  In particular, it is involved in the study of various decays and scatterings, beyond the SM \cite{georgi2007another,bhattacharyya2007unraveling,chen2009constraints,wei2009interpretation,mdah,jplee,abol,mden}.  Also many papers study the presence of unparticle in $CP$ violation, such as \cite{luo.,li2007unparticle,chen2007unparticle,mohanta2007possible,zwicky2008unparticles,huang2008direct,chen2008flavors,chen2010charge,ren2011large}. The effect of unparticle physics on the mixing of $B^0 - \overline{B^0}$ and $D^0 - \overline{D^0}$ have been considered in \cite{luo.} and \cite{li2007unparticle}, respectively. In Ref. \cite{chen2007unparticle} authors have been found that the phases in unparticle propagators have a great impact on $CP$ violation. Also, authors of Ref. \cite{huang2008direct}
  found that the direct $CP$ violation in the  $ B \rightarrow l \nu $ decay, which is zero in SM, can show up due to the $CP$ conserving phase
 intrinsic in unparticle physics. The effect of unparticle physics on $CP$ violation for decay $B^+ \rightarrow \tau^+ \nu$ have been also considered by Zwicky \cite{zwicky2008unparticles}.

In our discussion, unparticle physics contributes to one of the two amplitudes which are necessary for $CP$ violation.  Here, we first review the $CP$ asymmetries for $ D^0 \rightarrow K^- \pi^+ $ and $ D^+ \rightarrow K_s^0 \pi^+ $ decays in the SM and consider their reported values from various experiments. The first decay is Cabibbo-favored (CF). The second is a combination of $ D^+ \rightarrow \overline{K^0} \pi^+ $ and $ D^+ \rightarrow K^0 \pi^+ $ which are , respectively, CF and Doubly-Cabibbo-Suppressed (DCS) that is negligible. In the SM, for CF decays of D mesons there is only one amplitude with neither conserved-, nor violated-phase, so there is no predicted $CP$ violation for such decays. Here, we implement unparticle stuff to contribute as a second amplitude which can give us both conserved- and violated-phase differences. Using these decays, we try to constrain the
relevant parameter space of unparticle physics.

 We organized the paper as follows: in Sec. \ref{2} we  study the $ D^0 \rightarrow K^- \pi^+ $ and $ D^+ \rightarrow K_s^0 \pi^+ $ decays in the SM briefly and review the various experimental works on them. Then the unparticle effects for these decays is considered in Sec. \ref{3}. In the last section we conclude our results.

 \section{$ D^0 \rightarrow K^- \pi^+ $ and $ D^+ \rightarrow K_s^0 \pi^+ $ decays in the Standard Model and experiment}\label{2}

\subsection{$ D^0 \rightarrow K^- \pi^+ $ decay}\label{s1}

\begin{figure}[ht]
	\centerline{\includegraphics[scale=0.35]{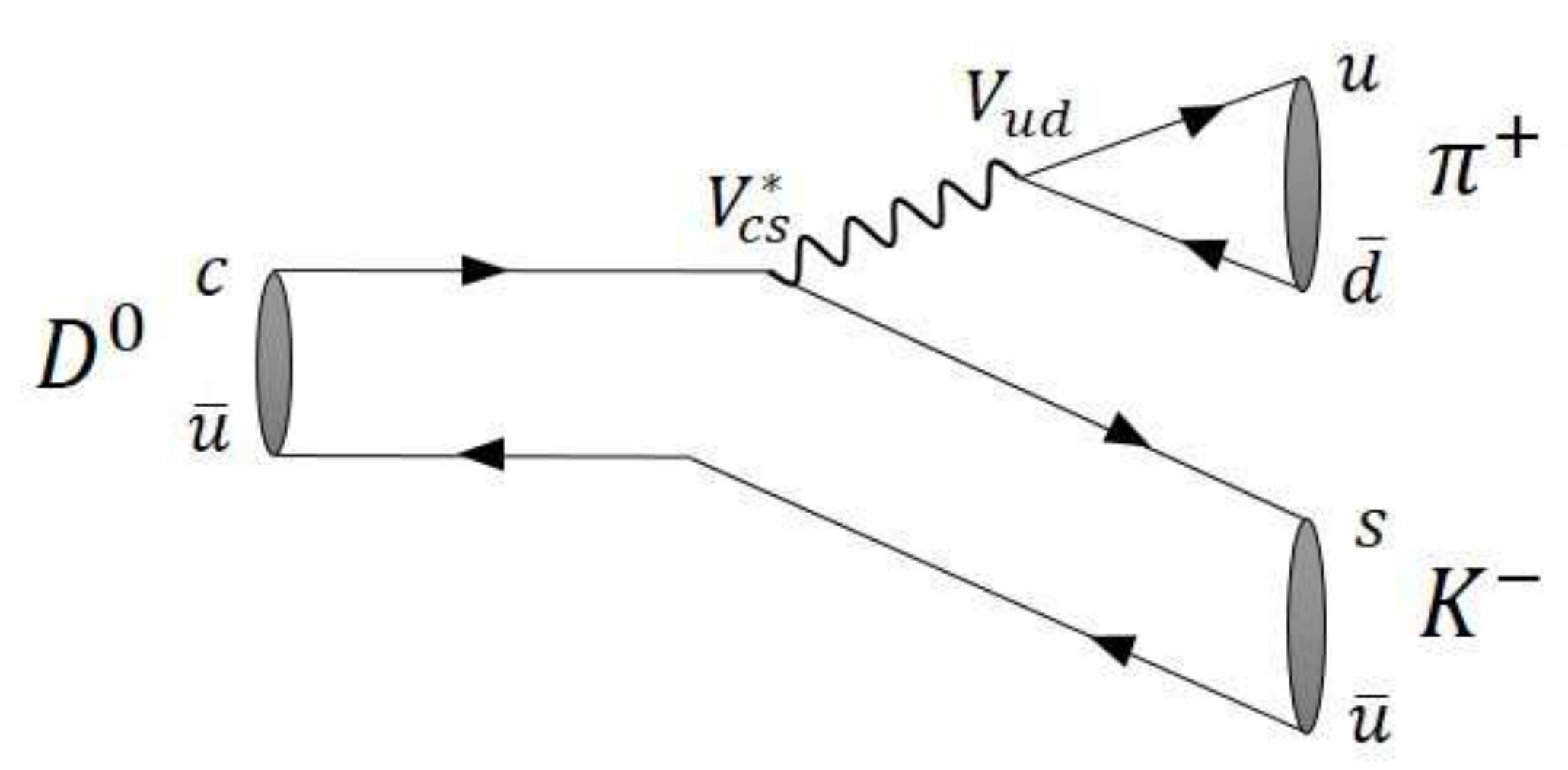}}
	\caption{Tree level $D^0 \rightarrow K^- \pi^+$ decay.}\label{f1}
\end{figure}

The main contributions to this decay are the tree level quark contribution,
exchange quark diagrams (box contribution) and color-suppressed quarks diagrams (di-penguins contribution). The tree level quark contribution is CF (see Fig. \ref{f1}).

 The direct $CP$ asymmetry is then \cite{delepine2013observation}
 \begin{eqnarray}
 A_{CP}^{D^0 \rightarrow K^- \pi^+}\equiv \dfrac{\Gamma (D^0 \rightarrow K^- \pi^+) - \Gamma (\overline{D^0} \rightarrow K^+ \pi^-) }{\Gamma (D^0 \rightarrow K^- \pi^+) + \Gamma (\overline{D^0} \rightarrow K^+ \pi^-)} = 1.4 \times 10^{-10} \, ,
 \end{eqnarray}
where $ \Gamma $ is the partial decay width. Due to the very smallness of this value, observation of a $CP$ violation for this decay can be a smoking gun of new physics.

Note that, the contribution of the indirect $CP$ violation for this decay is negligible.
The experimentally reported value for the $CP$ asymmetry in this decay, accepted by PDG, is $ A_{CP}= (0.3 \pm 0.7) \% $ \cite{pdg2016}.

 \subsection{$ D^+ \rightarrow K_s^0 \pi^+ $ decay }\label{s21}

  The first evidence of $CP$ violation in charmed particles reached after the FOCUS \cite{link2002search}, CLEO \cite{mendez2010measurements}, Belle \cite{ko2010search}, and BaBar \cite{del2011search} measurements for the decay $D^+ \rightarrow K_s^0 \pi^+$. The first world average for the $CP$ asymmetry of this decay was ($-0.54 \pm 0.14  \% $). This decay is performed through two steps; initially $ D^+ $ decays to $  K^0 $ or $ \overline{K^0} $, then $ K^0 \rightleftharpoons \overline{K^0} $ mixing occurs. For the $CP$ asymmetry of this decay we have
 \begin{eqnarray}
 A_{CP}^{D^+ \rightarrow K_s^0 \pi^+}\equiv \dfrac{\Gamma (D^+ \rightarrow K_s^0 \pi^+) - \Gamma (D^- \rightarrow K_s^0 \pi^-) }{\Gamma (D^+ \rightarrow K_s^0 \pi^+) + \Gamma (D^- \rightarrow K_s^0 \pi^-)} \, .
 \end{eqnarray}
  One can write, by a simple calculation,
 \begin{eqnarray}
 A_{CP}^{D^+ \rightarrow K_s^0 \pi^+}\approx A_{CP}^{\Delta C} + A_{CP}^{\rm{mixing}} \, ,
 \end{eqnarray}
where $ A_{CP}^{\Delta C} $ and $ A_{CP}^{\rm{mixing}}$ denote $CP$ asymmetries in the charm decay
($ \Delta C $) and in $ K^0 \rightleftharpoons \overline{K^0} $ mixing in the SM, respectively \cite{xing1995effect,bigi1995interference}.

\begin{figure}[ht]
 	\centerline{\includegraphics[scale=0.3]{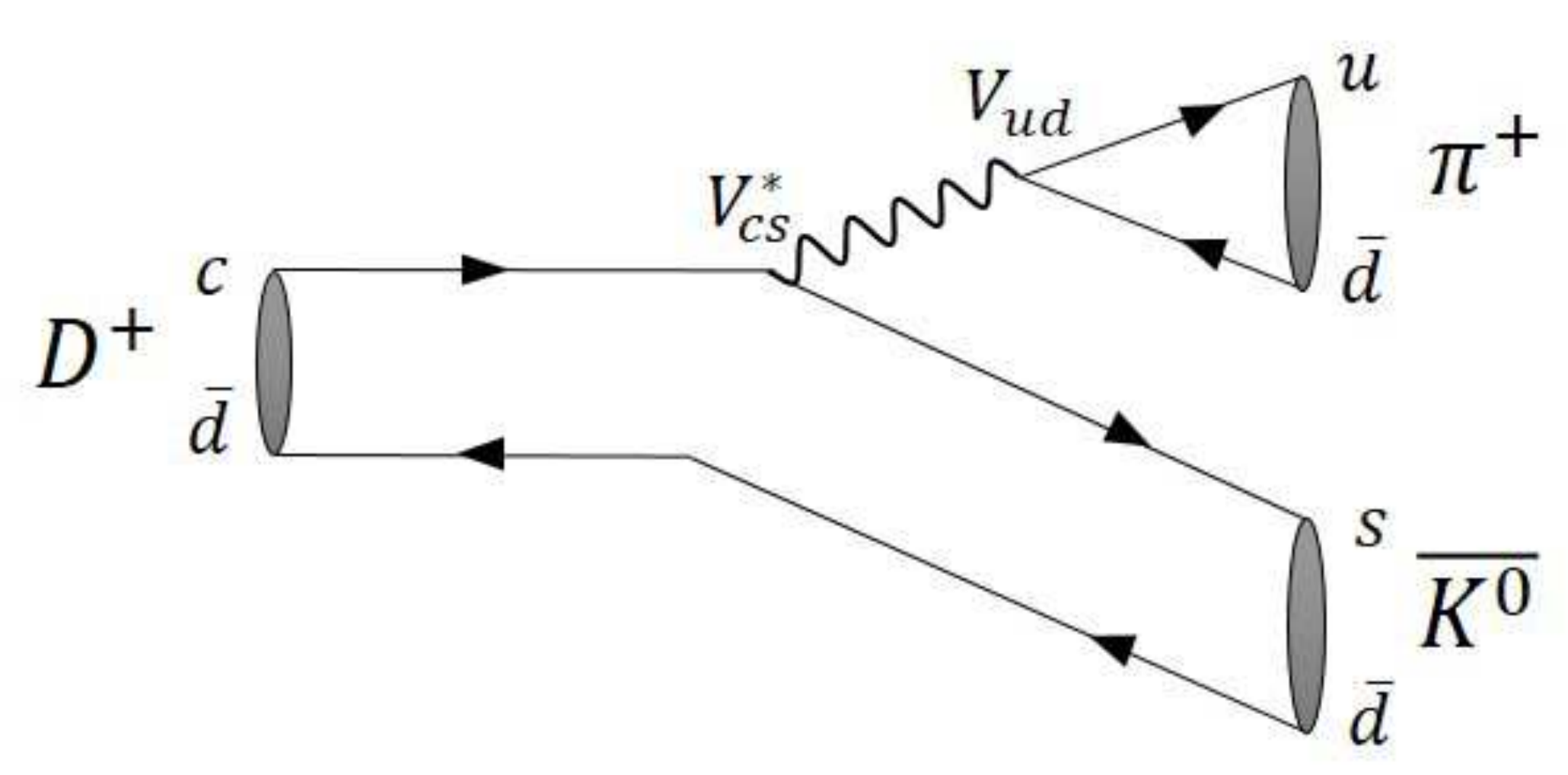}\includegraphics[scale=0.3]{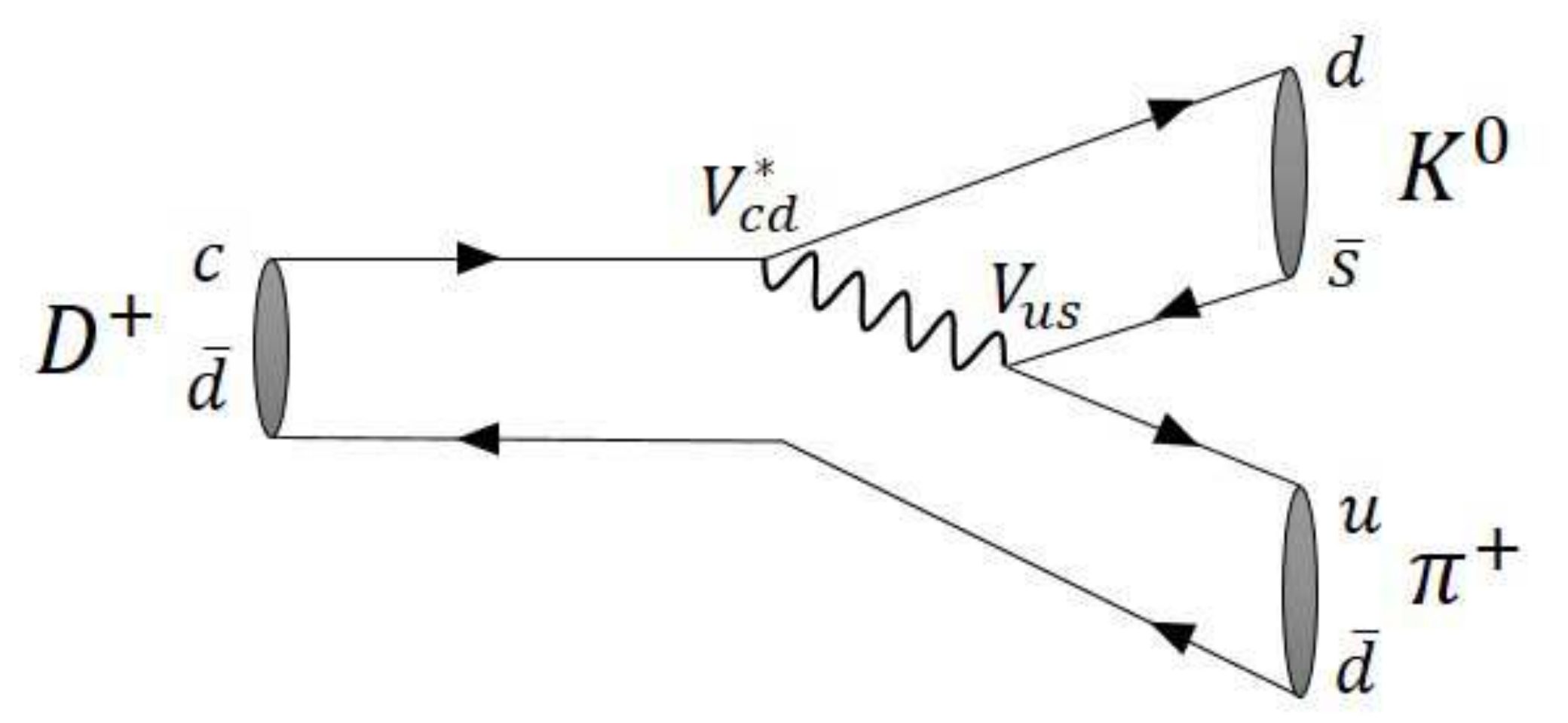}\hspace{-9.5cm}(a)\hspace{5.5cm}(b)\hspace{3cm}}
 	\caption{Tree level $D^+ \rightarrow \overline{K^0} \pi^+$  (left) and  $D^+ \rightarrow K^0 \pi^+$ (right). The first diagram is CF and the second is DCS.}\label{f4}
 \end{figure}

\begin{figure}[ht]
\centerline{\includegraphics[scale=0.35]{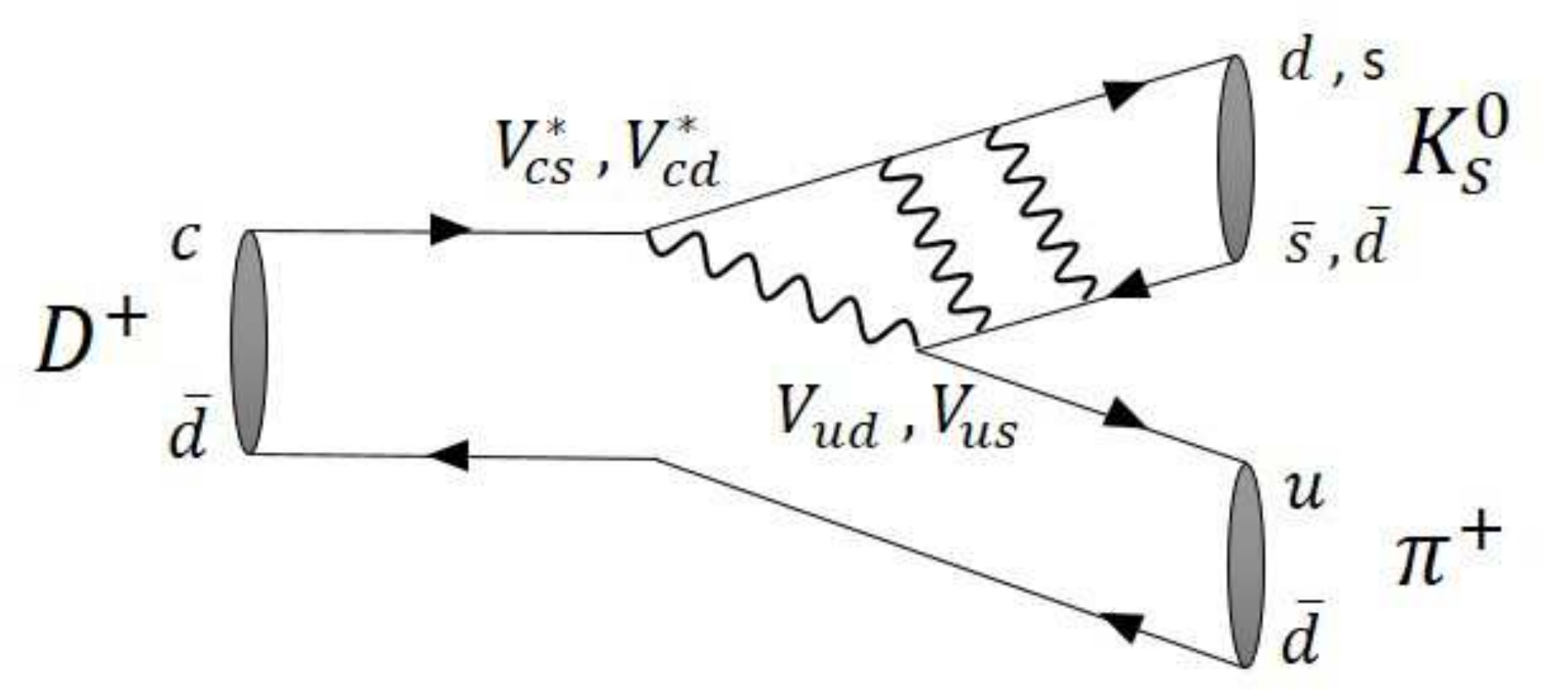}}
\caption{$D^+ \rightarrow K_s^0 \pi^+$ with both DCS and CF amplitudes.}\label{f5}
\end{figure}

 Amplitudes of two processes contribute to this decay; $ D^+ \rightarrow \overline{K^0} \pi^+ $ decay which is CF, Fig. \ref{f4}(a), and $ D^+ \rightarrow K^0 \pi^+ $ decay which is DCS, Fig. \ref{f4}(b).
 Mixing of $K^0$ and $\overline{K^0}$ in the final state leads to $K_s^0$. The combination of these two scenarios have been shown in Fig. (\ref{f5}).
On the other hand, for D decays, penguin diagrams, which we need for $CP$ violation as a second amplitude, contribute only to the singly Cabibbo suppressed (SCS) decays. Hence, focusing on CF decays, we have no $CP$ violation in charm sector. Consequently, all of the $CP$ asymmetry in $ D^+ \rightarrow K_s^0 \pi^+ $ must be due to $ K^0 \rightleftharpoons \overline{K^0} $ mixing, which is measured to be ($-0.332 \pm 0.006 $)$ \% $ from $ K_L^0 $ semileptonic decays ($ K_L^0 \rightarrow \pi^- l^+ \nu $) \cite{nakamura2010review}.
The $CP$ asymmetry values for this decay from various experiments are shown in Table \ref{table}. The new world average reported by PDG is $-0.41 \pm 0.09$ \cite{pdg}.
\begin{table}[ht]
\caption{$CP$ asymmetry for $ D^+ \rightarrow K_s^0 \pi^+ $ decay in different experiments \cite{ko2012evidence}}\label{table}
\begin{center}
\begin{tabular}{| c | c |}
\hline
Experiment & $ A_{CP}^{D^+ \rightarrow K_s^0 \pi^+} (\%) $ \\ \hline
FOCUS & -1.6 $ \pm $ 1.5 $ \pm $ 0.9 \\ \hline
CLEO & -1.3 $ \pm $ 0.7 $ \pm $ 0.3 \\ \hline
BaBar & -0.44 $ \pm $ 0.13 $ \pm $ 0.10 \\ \hline
Belle & -0.363 $ \pm $ 0.094 $ \pm $ 0.067 \\ \hline \hline
New world average & -0.41 $ \pm $ 0.09 \\ \hline
\end{tabular}
\end{center}
\end{table}
Therefore, comparing the mixing contribution reported from $ K_L^0 $ semileptonic decays, and the  new world average, we see about ($ -0.08\pm 0.09$)\%  of asymmetry difference. 
Consequently, the contribution of any possible NP, such as unparticle, in  $CP$ asymmetry of charm sector should lie in this interval. Hereby, we can constrain the parameter space of unpartcle stuff.

\section{$ D^0 \rightarrow K^- \pi^+ $ and $ D^+ \rightarrow K_s^0 \pi^+ $  with unparticle}\label{3}
In this section we first, briefly, review the  unparticle physics  which  is a new scale invariant sector introduced firstly by Georgi \cite{georgi2007unparticle}.
 The propagator of a scalar (vector) unparticle  $O_{\cal{U}}^{(\mu)}$, is 
 \begin{eqnarray}
 \int d^{4}xe^{ip.x}\langle0\vert T\left[O^{(\mu)}_{\cal{U}}(x)O^{(\nu)}_{\cal{U}}(0)\right]\vert0\rangle=\Delta^{S(V)}_{\cal{U}}(p^{2})e^{-i\phi_{\cal{U}}}\, ,
 \end{eqnarray}
where
\begin{eqnarray}
 &\Delta^{S}_{\cal{U}}(p^{2})=\frac{A_{d_{\cal{U}}}}{2\sin (d_{\cal{U}}\pi)}\frac{1}{(p^{2}+i\epsilon)^{2-d_{\cal{U}}}}\, ,
 \nonumber\\
 \nonumber \\
 &\Delta^{V}_{\cal{U}}(p^{2})=\frac{A_{d_{\cal{U}}}}{2\sin (d_{\cal{U}}\pi)}\frac{-g^{\mu\nu}+p^{\mu}p^{\nu}/p^{2}}{(p^{2}+i\epsilon)^{2-d_{\cal{U}}}}\,,
 \end{eqnarray}
 are the scaler and vector propagators, respectively. Here, $d_{\cal{U}}$ is the unparticle dimension, $\phi_{\cal{U}}=(2-d_{\cal{U}})\pi$ and

 \begin{eqnarray}
 A_{d_{\cal{U}}}=\frac{16\pi^{5/2}}{(2\pi)^{2d_{\cal{U}}}}\frac{\Gamma(d_{\cal{U}}+1/2)}{\Gamma(d_{\cal{U}}-1)\Gamma(2d_{\cal{U}})}\, .
 \end{eqnarray}
 Then, the unparticle couplings with quarks will be given by the following effective Lagrangian:
  \begin{eqnarray}
 \mathcal{L} =  \frac{c^{q^{\prime} q}_{V}}{\Lambda_{\cal{U}}^{d_{\cal{U}}-1}}\bar{q'} \gamma_{\mu}(1-\gamma_{5})q O^{\mu}_{\cal{U}}+\frac{c^{q^{\prime} q}_{S}}{\Lambda_{\cal{U}}^{d_{\cal{U}}}}\bar{q'} \gamma_{\mu}(1-\gamma_{5})q \partial_{\mu}O_{\cal{U}} + \rm{H.C.} \, ,
 \end{eqnarray}
 where $ c^{q^{\prime} q}_{S,V} $  are dimensionless parameters and $\Lambda_{\cal{U}}$ is an energy scale  in which unparticles will appear. The first (second) term in this Lgrangian is related to the vector (scalar) unparticle. Unparticle with scale dimension $ d_{\cal{U}} $ treats as nonintegral number $ d_{\cal{U}} $ of invisible  massless particle. 

 \subsection{$ D^0 \rightarrow K^- \pi^+ $ decay with tree level unparticle amplitude}\label{s2}

\begin{figure}[ht]
 	\centerline{\includegraphics[scale=0.35]{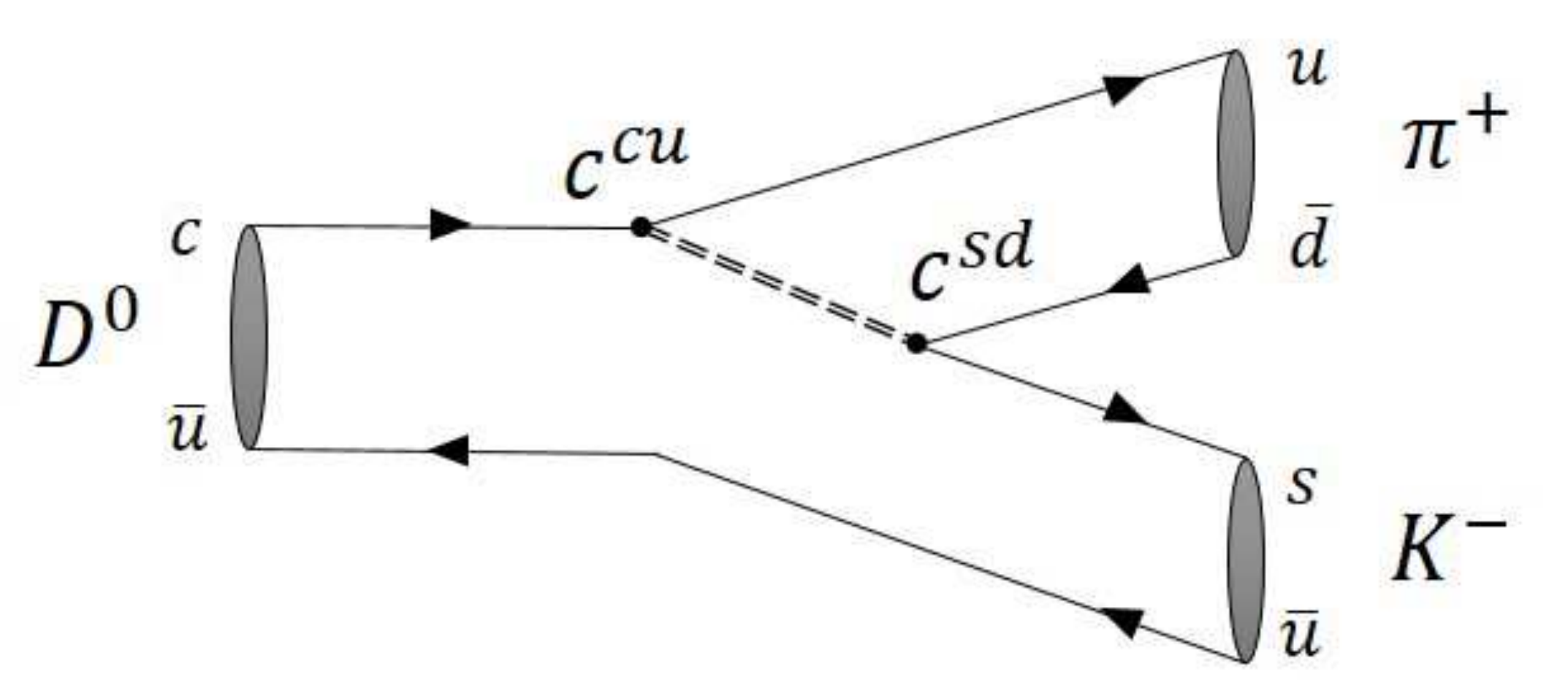}}
 	\caption{$D^0 \rightarrow K^- \pi^+$  with unparticle.}\label{f2}
 \end{figure}

Now, we investigate the $ D^0 \rightarrow K^- \pi^+ $ decay with unparticle. As mentioned in Sec. \ref{s1}, in the SM, this decay has only a CF amplitude at tree level with no penguin diagram. The unparticle diagram for this decay is shown in Fig. \ref{f2} (Here we consider an uncharged unparticle). This new amplitude can give us strong ($CP$-conserved) and weak ($CP$-violated) phase differences needed for $CP$ violation. The total amplitude now becomes
 \begin{equation}
 A^{K^- \pi^+}_{\rm{total}} = A^{K^- \pi^+}_{\rm{SM}} + A^{K^- \pi^+}_{\cal{U}} =A_{\rm{SM}}^{K^- \pi^+} \left(1+r_{_{K^-\pi^+}} \ e^{-i\phi_{\cal{U}}} e^{-i\phi_{W}} \right) \,,
 \end{equation}
  where $\phi_{W}$ and $\phi_{\cal{U}} = \left( d_{\cal{U}}-2 \right) \pi $ are $CP$-violated and $CP$-conserved phase differences, respectively (knowing that the SM phases are zero). Here, $A_{\rm{SM}}^{K^- \pi^+}$ is the SM amplitude \cite{zwicky2008unparticles}
 \begin{equation}
 A_{\rm{SM}}^{K^- \pi^+} = \dfrac{G_F}{\sqrt{2}}  V_{cs}^*V_{ud}  {\cal F}\,,
 \end{equation}
 and $r_{_{K^-\pi^+}}$ is the ratio of unparticle and SM amplitudes,
 \begin{equation}
 r_{K^-\pi^+} =\frac{8}{g^{2}a_{1}N_{c}}\frac{|c_{V}^{cu}c_{V}^{sd}|}{\vert V_{cs}^{*}V_{ud}\vert}\frac{A_{d_{\cal{U}}}}{2\sin (d_{\cal{U}}\pi)}\frac{m_{W}^{2}}{p^{2}}\left(\frac{p^{2}}{\Lambda_{\cal{U}}^2}\right)^{d_{\cal{U}}-1}\,,
 \end{equation}
where $\cal F$ is a function which depends on the meson mass and QCD detail, that finally removed in Eq. (\ref{e2}). Here, $a_1 = C_2+C_1/N_c$ is the effective Wilson coefficient
\cite{buchalla}, $ N_c$ is the color number and $ p^2 \sim m_D \bar{\Lambda}$ with  $\bar{\Lambda} =
 m_D-m_{\rm{c}}$. We can ignore the 
 scalar unparticle contributions, since they  are suppressed by $m_D^2
/\Lambda_{\cal U}^2$. Therefore, the Eq. (\ref{e2}) becomes
 \begin{equation}\label{aaa}
 A^{K^- \pi^+}_{CP} = \dfrac{2 \ r_{K^-\pi^+} \sin({d_{\cal{U}}\pi}) \sin{\phi_{W}}}{1+r_{K^-\pi^+}^2+2 r_{K^-\pi^+} \cos{d_{\cal{U}}\pi} \cos{\phi_{W}}} \,.
 \end{equation}
  Note that, in this case, as we mentioned before, up to order $\lambda^4$ in the Wolfenstein CKM matrix there is no CKM weak phase and here $\phi_{W}$ ($CP$-violated phase) comes completely from the complex valued nature of unparticle couplings.

 Here, we try to illustrate the role of the various parameters in some figures in such a way $ A_{CP}^{D^0 \rightarrow K^- \pi^+}=(0.3 \pm 0.7) \%$. There are, in principle, four independent parameters: the unparticle scale $\Lambda_{\cal U}$, the scaling dimension $d_{\cal U}$, the net resultant phase of the coupling constants $\phi_W$, the absolute value of couplings product, $|c_{V}^{cu}c_{V}^{sd}|$. We fix the scale of unpartcle $\Lambda_{\cal U}$ about 15 TeV, due to the recent energy achievement in LHC. The dependence of $A_{CP}$ to $|c_{V}^{cu}c_{V}^{sd}|$ is also trivial (linear), in the regime where we expect the perturbation works and also  corrections to the SM are small\footnote{The upper bound of $ c_{V}^{cu} $ have been fixed  from $ D^0 \rightleftharpoons \overline{D^0} $ mixing at $|c_{V}^{cu}|< 5\times 10^{-4}$  for $ d_{\cal{U}}=3/2 $\cite{li2007unparticle}.}. Therefore the main parameters which may play important roles are $\phi_W$ and $d_{\cal U}$.

\begin{figure}[ht]
 	\centerline{\includegraphics[scale=0.5]{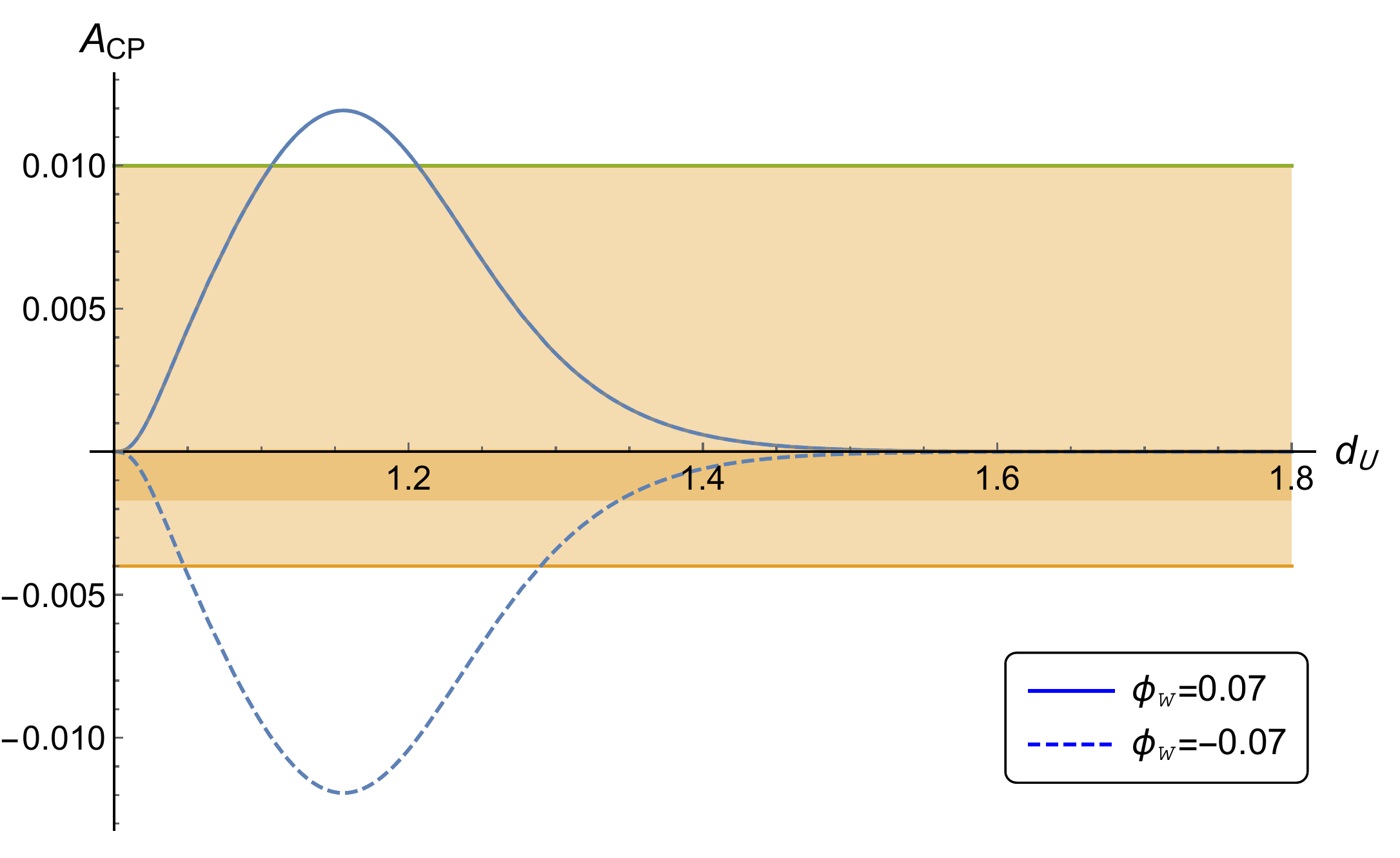}}
 	\caption{$A_{CP}$ (due to the charm sector) in terms of $d_{\cal{U}}$, for $\Lambda_{\cal U}=15$ TeV, $|c^{sd}_V c^{cu}_V|  \approx 10^{-5}$,  $\phi_W=0.07$ (solid line) and $\phi_W=-0.07$ (dashed line). The dark region is related to the experimental bound for $CP$ asymmetry for $D^0 \rightarrow K^- \pi^+$ decay and the darker one for $ D^+ \rightarrow K_s^0 \pi^+ $ decay. }\label{fn5}
 \end{figure}

\begin{figure}[th]
\begin{center}$
	\begin{array}{ll}
	\includegraphics[scale=0.63]{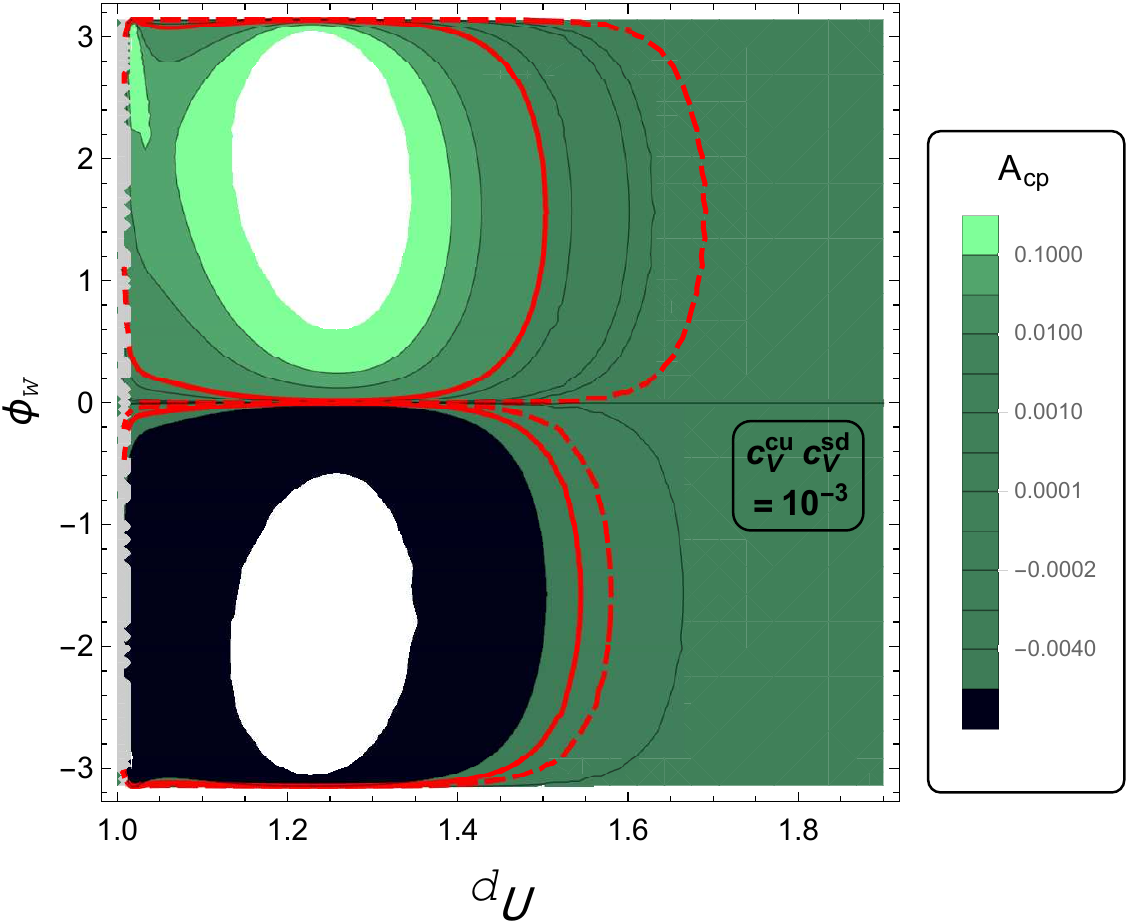}&\includegraphics[scale=0.63]{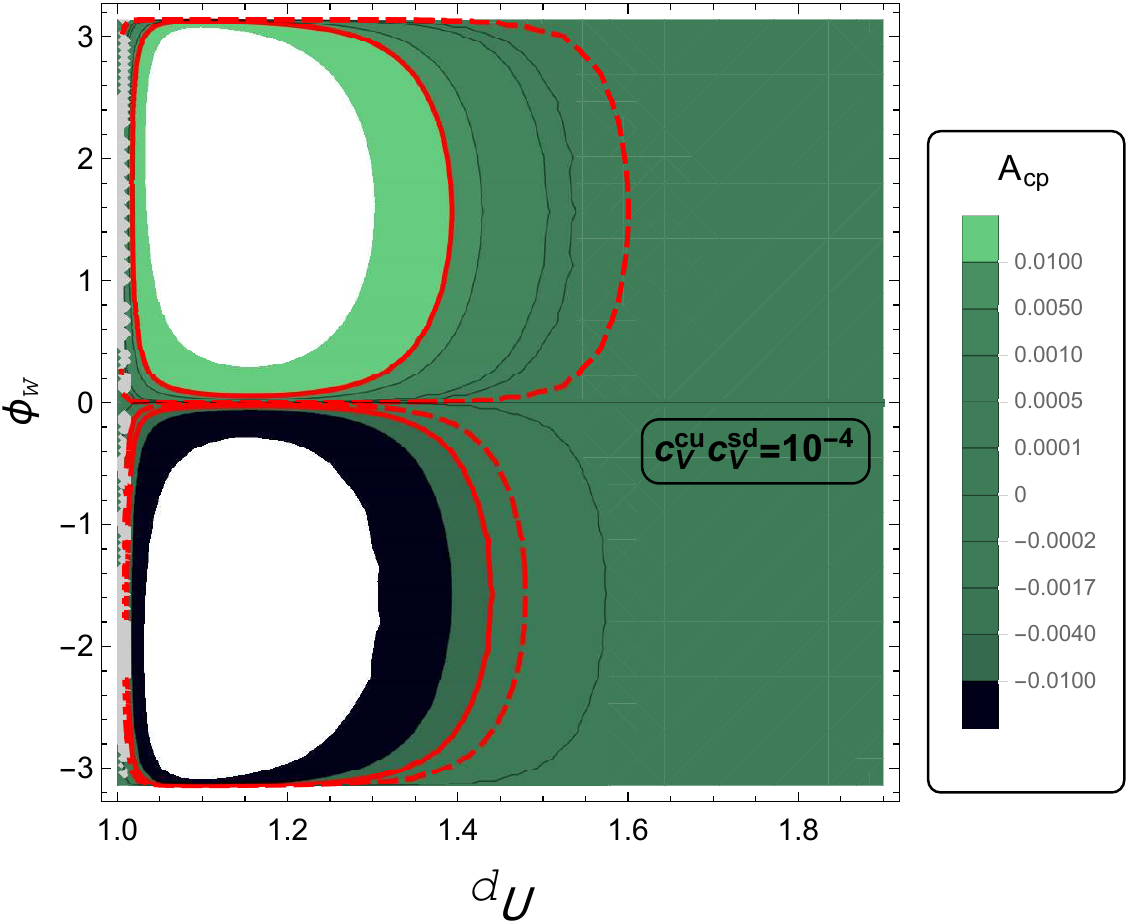} \\	\includegraphics[scale=0.63]{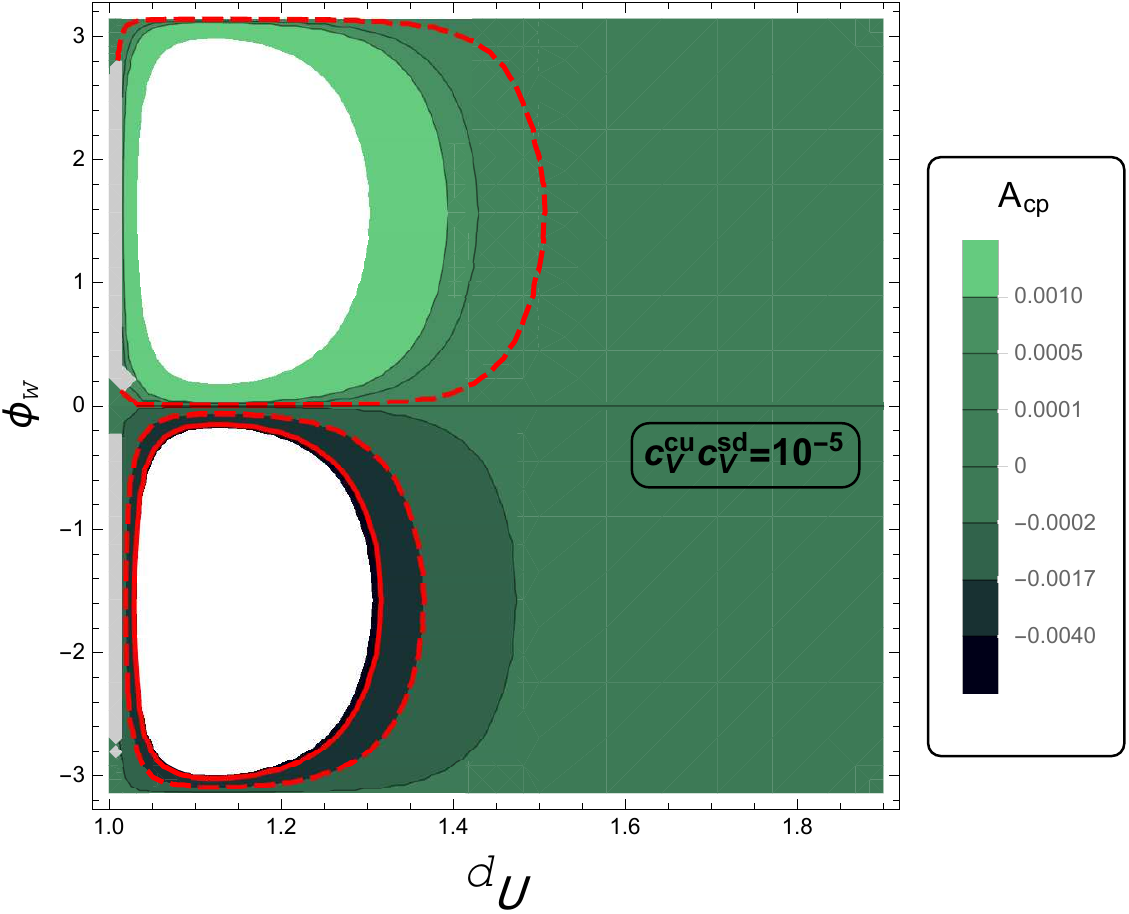}&\includegraphics[scale=0.63]{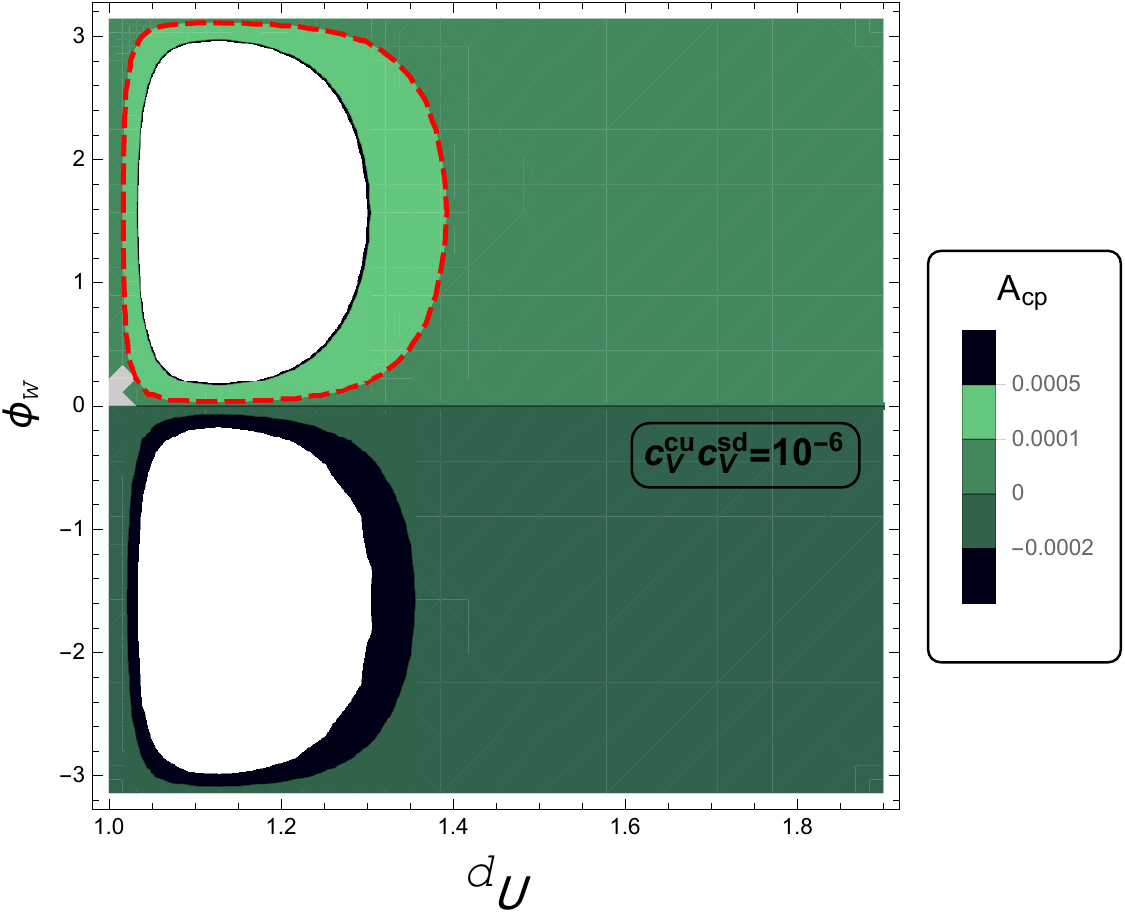} 
	\end{array}$
	\end{center}
	\caption{$A_{CP}$ (color) for $D^0 \rightarrow K^-\pi^+$ and also for charm sector of $D^+ \rightarrow K_s^0\pi^+$ with respect to $\phi_W$ and $d_{\cal U}$ for different values of $|c^{sd}_V c^{cu}_V|$. The red solid (dashed) lines denotes the recent bounds on $CP$ asymmetry for the first (second) decay. } \label{figcp}
\end{figure}

  In Fig. \ref{fn5}, we have plotted $ A_{CP} $ in terms of $d_{\cal{U}}$ for two different  $\phi_W=\pm 0.07$ . In this figure, the allowed region has been colored and we have fixed  $|c^{sd}_Vc^{cu}_V| \approx 10^{-5}$. This figure shows that for some regions, around $d_{\cal U}=1.15$, $A_{CP}$ is larger than all allowed values. 
  Since the weak and strong phases play important roles in $CP$ violation, we try to study the related parameter space for some various values of product $|c^{sd}_V c^{cu}_V| $ through contour plots of Fig. \ref{figcp}. In these contour plots the vertical axes shows the weak phase, and the horizontal axes is devoted to dimension of the unparticle $d_{\cal U}$ which determines the strong phase. Here, the color shows the $A_{CP}$.
  
   As a result of this figure, in $\Lambda_{\cal{U}}=15 $ TeV, there exist some regions which are excluded by this process for $|c^{sd}_V c^{cu}_V| \sim 10^{-5}$, while in the case of $|c^{sd}_V c^{cu}_V| \sim 10^{-6}$, or weaker couplings, the whole of parameter space is allowed. In other words, for $|c^{sd}_V c^{cu}_V| \approx 10^{-5}$ as well as stronger couplings, the unparticle physics contribution can be explored by experimental test which is more accurate than the recent data while for the values less than about $10^{-5}$, it is far from recent precisions and is not testable.

 \begin{figure}[ht]
 	\centerline{\includegraphics[scale=0.35]{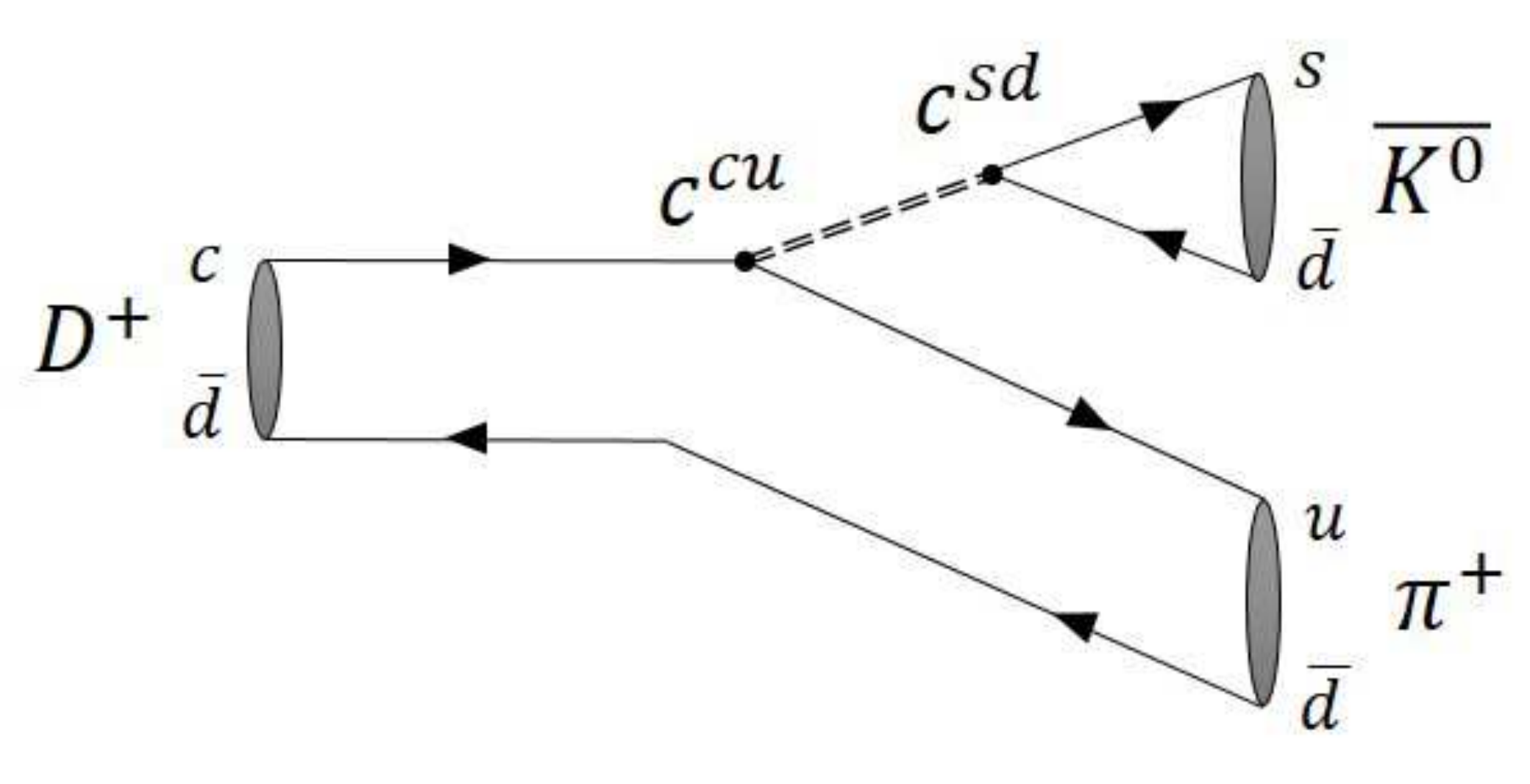}}
 	\caption{$D^+ \rightarrow \overline{K^0} \pi^+$ decay with unparticle.}\label{f6}
 \end{figure}
 
 \begin{figure}[ht]
 	\centerline{\includegraphics[scale=0.35]{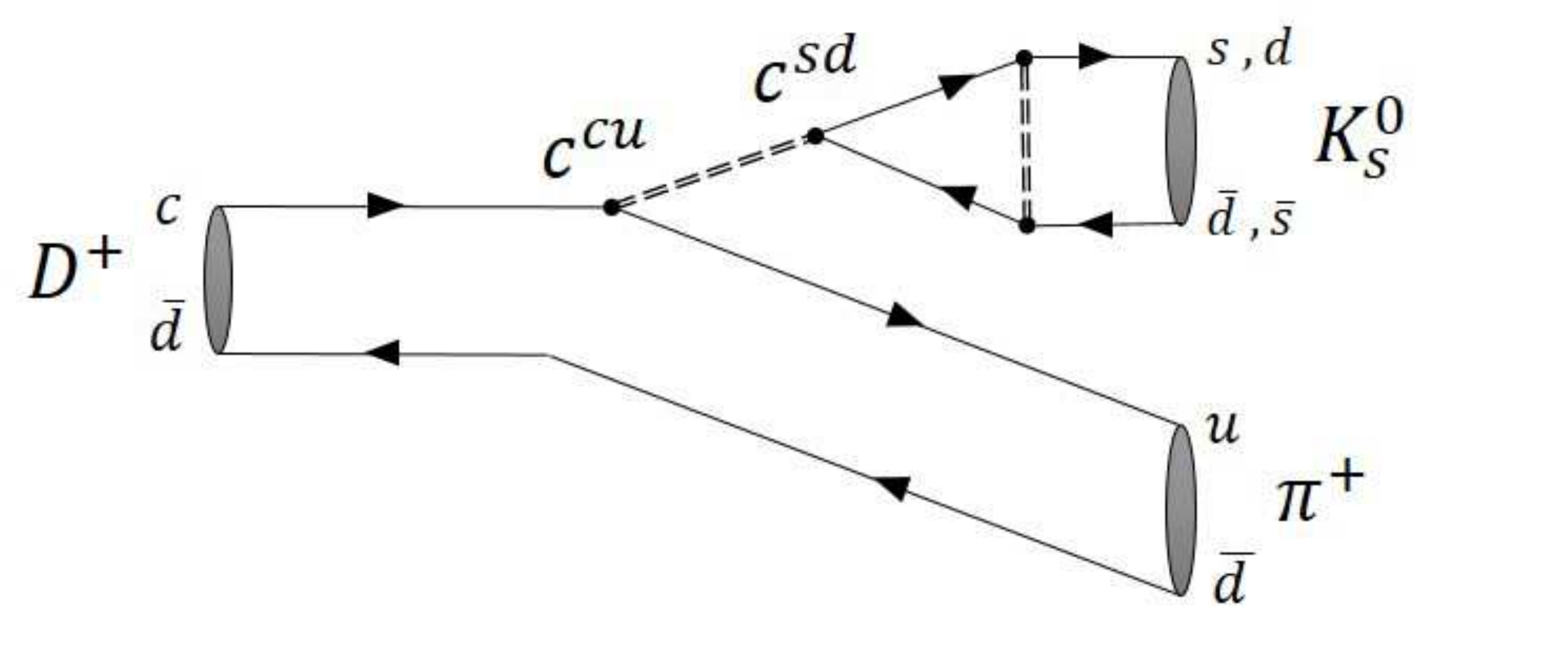}}
 	\caption{$D^+ \rightarrow K_s^0 \pi^+$ decay with unparticle.}\label{f7}
 \end{figure}

 \subsection{$ D^+ \rightarrow K_s^0 \pi^+ $ decay with tree level  unparticle amplitude }\label{s3}

Here in this section, we apply unparticle theory as a second amplitude to explain the ($ -0.08\pm 0.09$)\% $CP$ asymmetry related to the charm sector. As mentioned before, this value is due to the difference between the  world average $CP$ asymmetry for $ D^+ \rightarrow K_s^0 \pi^+ $ decay and the corresponding value for $ K^0 \rightleftharpoons \overline{K^0} $ mixing ($ A_{CP}^{\rm{mixing}} $).

 Again, for this decay we have no penguin diagram and in tree level it has both CF and DCS amplitudes which we neglect DCS one \cite{grossman2012cp}. Unparticle stuff can give a diagram which leads to the strong and weak phase differences between  two amplitudes (corresponding to SM and unparticle).  One could see the diagram of $ D^+ \rightarrow \overline{K^0} \pi^+ $ and   $ D^+ \rightarrow K_s^0 \pi^+ $ with unparticle in Figs. \ref{f6} and  \ref{f7}, respectively.

  This decay is the same as $ D^0 \rightarrow K^- \pi^+ $, if one changes the observer quark $ \bar{u} $ to $ \bar{d} $ and also adds a $ K^0 \rightleftharpoons \overline{K^0} $ mixing in final state. To write the total $A_{CP}$ we note that,  we are seeking for a $CP$ asymmetry in addition to the contribution of SM mixing, as mentioned before. Moreover, the unparticle contribution in $  K^0 \rightleftharpoons \overline{K^0}$ mixing is negligible  \cite{chen2009constraints}. Therefore, the total amplitude becomes

  \begin{eqnarray}
 A_{\rm{total}} &=& A_{\rm{SM}}^{\overline{K^0} \pi^+ } + A_{\cal{U}}^{\overline{K^0} \pi^+ }\nonumber\\
&=&A_{\rm{SM}}^{\overline{K^0} \pi^+ } \left(1+r_{_{\overline{K^0}\pi^+}} \ e^{-i\phi_{\cal{U}}} e^{-i\phi_{W}} \right) \,,
 \end{eqnarray}
 where
 \begin{equation}
 A_{\rm{SM}}^{ \overline{K^0} \pi^+} = \dfrac{G_F}{\sqrt{2}}  V_{cs}^*V_{ud} {\cal F'}\,,
 \end{equation}
 and
 \begin{equation}
 r_{\overline{K^0}\pi^+} =\frac{8}{g^{2}a_{1}N_{c}}\frac{|c_{V}^{cu}c_{V}^{sd}|}{\vert V_{cs}^{*}V_{ud}\vert}\frac{A_{d_{\cal{U}}}}{2\sin (d_{\cal{U}}\pi)}\frac{m_{W}^{2}}{p^{2}}\left(\frac{p^{2}}{\Lambda_{\cal{U}}^2}\right)^{d_{\cal{U}}-1}\,,
\end{equation}
where $\cal F'$ is defined similar to $\cal F$. Consequently,  the direct $CP$ asymmetry becomes

 \begin{equation}
 A_{CP} \approx A_{CP}^{\rm{mixing}} + \dfrac{2 \ r_{\overline{K^0}\pi^+} \sin({d_{\cal{U}}\pi}) \sin{\phi_{W}}}{1+r_{\overline{K^0}\pi^+}^2+2 r_{\overline{K^0}\pi^+} \cos{d_{\cal{U}}\pi} \cos{\phi_{W}}} \,.
 \end{equation}
As it is obvious from the above equation, the second term in the right-hand side is exactly the same as Eq. (\ref{aaa}). Therefore, our general phenomenological discussion do not alter, however, here we should be careful about the allowed regions in parameter space (see Figs. \ref{fn5} and \ref{figcp}). In particular, according to Fig. \ref{figcp}, for $|c^{sd}_V c^{cu}_V| \sim 10^{-6}$, while the whole region is allowed in the case of first process, some region with positive value of $A_{\rm CP}$ is excluded by the recent process.

\section{Conclusions}

The new world averages for $CP$ violation in $ D^0 \rightarrow K^- \pi^+ $ and $ D^+ \rightarrow K_s^0 \pi^+ $ decays reported by PDG are ($0.3 \pm  0.7$) $\%$ and ($-0.41 \pm  0.09$) $\%$ respectively. In $ D^+ \rightarrow K_s^0 \pi^+ $, the value ($-0.332 \pm 0.006 $) $\%$ is due to the mixing of $ K^0 $ and $ \overline{K^0} $ mesons in the final state. Subtracting this contribution, one can conclude that any possible NP gives, at most, a $CP$ asymmetry in the interval ($-0.08 \pm  0.09$) $\%$.  Interaction between a scale invariant sector, called unparticle by Georgi \cite{georgi2007unparticle}, and the SM fields, as a NP, can induce a $CP$ asymmetry \cite{chen2007unparticle}.

 In this paper, we have studied the unparticle induced $CP$ asymmetry in both processes  $ D^0 \rightarrow K^- \pi^+ $ and $ D^+ \rightarrow K_s^0 \pi^+ $ decays. More explicitly, both phase differences (weak and strong), needed for $CP$ violation, come from unparticle diagrams. Note that, these two decays are CF (in charm sector), which has no predicted $CP$ asymmetry in the SM at tree level. Here, in addition to the scale of unparticle physics ${\Lambda_{\cal U}}$, three important parameters play role; the net resultant weak phase of unparticle $\phi_W$, the dimension of unparticle $d_{\cal U} $ which determines the strong phase and the product of couplings $|c^{sd}_V c^{cu}_V| $. The $CP$ asymmetry with respect to the $d_{\cal U} $ for a fixed value of $\phi_W$ and $|c^{sd}_V c^{cu}_V| $ is plotted for  ${\Lambda_{\cal U}=15}$ TeV in Fig. \ref{fn5}. With choosing $\phi_W=\pm0.2$ and  $|c^{sd}_V c^{cu}_V| =10^{-5}$, for instance, the absolute value of $CP$ asymmetry gets a maximum in $d_{\cal U}\sim 1.2$, in which the $CP$ asymmetry exceeds of experimental bounds.  We have also demonstrated the parameter space of this theory through some contour plots for ${\Lambda_{\cal U}=15}$ TeV and various values of $|c^{sd}_V c^{cu}_V| $. We see excluded regions in all selected $|c^{sd}_V c^{cu}_V| $, which  correspond to the pick regions of $CP$ asymmetry diagram.

\end{document}